# Predicting potential treatments for complex diseases based on miRNA and tissue specificity

**Liang Yu***

School of Computer Science and Technology, Xidian University, Xi'an, 710071, P.R.China

lyu@xidian.edu.cn

*Corresponding author

**Jin Zhao**

School of Computer Science and Technology, Xidian University, Xi'an, 710071, P.R.China

zhaojing-2108@qq.com

**Lin Gao**

School of Computer Science and Technology, Xidian University, Xi'an, 710071, P.R.China

lgao@mail.xidian.edu.cn

# Abstract

Network-based computational method, with the emphasis on biomolecular interactions and biological data integration, has succeeded in drug development and created new directions, such as drug repositioning and drug combination. Drug repositioning, that is finding new uses for existing drugs to treat more patients, offers time, cost and efficiency benefits in drug development, especially when in silico techniques are used. microRNAs (miRNAs) play important roles in multiple biological processes and have attracted much scientific attention recently. Moreover, cumulative studies demonstrate that the mature miRNAs as well as their precursors can be targeted by small molecular drugs. At the same time, human diseases result from the disordered interplay of tissue- and cell lineage-specific processes. However, few computational researches predict drug-disease potential relationships based on miRNA data and tissue specificity. Therefore, based on miRNA data and the tissue specificity of diseases, we propose a new method named as *miTS* to predict the potential treatments for diseases. Firstly, based on miRNAs data, target genes and information of FDA (Food and Drug Administration) approved drugs, we evaluate the relationships between miRNAs and drugs in the tissue-specific PPI (protein-protein) network. Then, we construct a tripartite network: drug-miRNA-disease Finally, we obtain the potential drug-disease associations based on the tripartite network. In this paper, we take breast cancer as case study and focus on the top-30 predicted drugs. 25 of them (83.3%) are found having known connections with breast cancer in CTD (Comparative Toxicogenomics Database) benchmark and the other 5 drugs are potential drugs for breast cancer. We further evaluate the 5 newly predicted drugs from clinical records, literature mining, KEGG pathways enrichment analysis and overlapping genes between enriched pathways. For each of the 5 new drugs, strongly supported evidences can be found in three or more aspects. In particular, Regorafenib (DB08896) has 15 overlapping KEGG pathways with breast cancer and their p-values are all very small. In addition, whether in the literature curation or clinical validation, Regorafenib has a strong correlation with breast cancer. All the facts show that Regorafenib is likely to be a truly effective drug, worthy of our further study. It further follows that our method *miTS* is effective and practical for predicting new drug indications, which will provide potential values for treatments of complex diseases.

# Introduction

The identification of therapeutic approaches for the treatment of cancer is an arduous, costly, and often inefficient process. By conservative estimates, it now takes over 15 years and $800 million to $1 billion to bring a new drug to market[1]. Drug repositioning, which is the discovery of new indications for existing drugs, is an increasingly attractive mode of therapeutic discovery. A repositioned drug does not need the initial six to nine years required for the development of new drugs, but instead goes directly to preclinical testing and clinical trials, thus reducing risk and costs[2]. Repositioning drugs has been implemented in several ways. One of the well-known examples is sildenafil citrate, which was repositioned from a hypertension drug to a therapy for erectile dysfunction[3]. Drugs treat diseases by targeting the proteins related to the phenotypes arising from the disease. However, drug development does not accurately follow the "one gene, one drug, one disease" principle, which has been challenged in many cases[4] and the traditional drug repositioning methods by accident makes it hard to satisfy medical needs by successfully repositioning a large number of existing drugs. Computational methods are able to solve this question by high-level integration of available biological data and elucidation of unknown mechanisms.

In recent years, systems biology continues to make important progress to solve fundamental problems in biology and leading to practical applications in medicine and drug discovery[5]. Network-based computational systems biology emphasizes the interactions among biomolecules and highlights the network concept. Typically, a network comprises a set of nodes and edges, and is described by graph theory in a mathematical manner[6]. A node can be a biological molecule, for example, gene, RNA, protein, metabolite, and pathway. A node can also be at the phenotype level such as disease and drug. An edge can represent the complex interaction between two nodes such as protein-protein interaction, drug-disease therapeutic relationship, drug-protein target relationship, and so on. The accumulation of different high-throughput biology data, such as gene expression data, miRNA expression data and drug-target data, has made the reconstruction of biomolecular and cellular networks possible. Cheng et al. built a bipartite graph composed of the approved drugs and proteins linked by drug target binary associations, and relied on a supervised network-based inference method to predict drug-target interactions[7]. Chen et al. constructed a general heterogeneous network which comprised drugs and proteins linked by protein-protein sequence similarity, drug-drug chemical similarity, and the known drug-target interaction[8]. Yeh et al.[9] developed a network flow approach for identifying potential target proteins, which have a strong influence on disease genes in the context of biomolecular networks. The biomolecular networks are weighted by degree of co-expression of interacting protein pair.

More recently, many studies have demonstrated that drugs can regulate microRNA (miRNAs) expression and mature miRNAs as well as their precursors can be targeted by small molecular drugs[10,11,12,13]. For example, Miravirsen (SPC3649) is the first miRNA-targeted drug in clinical trials, which can successfully inhibit miR-122 expression that is required by hepatitis C virus replication[14]. The expression levels of 32 miRNAs (significant up-regulation of 22 miRNAs and down-regulation of 10 miRNAs) were changed after the treatment of trichostatin A in human breast cancer cell lines [15]. miRNAs are non-coding small RNAs (~23 nucleotides) that downregulate gene expression at the post transcriptional level by inhibiting translation or initiating mRNA degradation and are dysregulated in most of human cancers[16]. Increasingly evidences have demonstrated that miRNAs play significant roles in many important biological processes, such as cell growth[17], cellular signaling[18], tissue development[19] and disease process[20]. Although only approximately 2000 miRNAs exist in humans, they regulate 30% of all genes. miRNAs have been identified to play a crucial role in various human disease, especially in cancers. Therefore, targeting miRNAs with drugs will provide a new type of therapy for complex diseases[21] and a new direction for drug repositioning. However, few computational researches predict drug-disease relationships based on miRNA data. Moreover, many genes with tissue-specific expression and function are expected to underlie many human diseases[22,23].

Therefore, in this study, we propose a new method based on miRNA data and tissue specificity of diseases, named as *miTS*, to predict potential drugs for diseases. The framework of *miTS* is shown in Figure 1. Firstly, we download miRNA expression data of diseases from TCGA[24], miRNA-target gene relationship data from three experimentally validated databases: miRecords[25], miRTarbase[26] and TarBase[27], and the drug-target gene data from Drugbank[28] and KEGG[29]. Secondly, we select differentially expressed miRNAs of diseases based on a threshold and preprocess the target information of FDA approved drugs. Thirdly, we evaluate the relationships between miRNAs and drugs in the tissue-specific PPI network. And then, we construct a tripartite network: drug-miRNA-disease. Finally, we obtain the potential drug-disease associations based on the tripartite network. In this paper, we take breast cancer as case study and evaluate the results from CTD benchmark, clinical records, literature mining, KEGG pathways enrichment analysis and overlapping genes between enriched pathways. In the top-30 drugs, we find 5 new drugs for breast cancer. In particular, Regorafenib (DB08896) has 15 overlapping KEGG pathways with breast cancer and their p-values are all very small. In addition, whether in the literature curation or clinical validation, Regorafenib has a strong correlation with breast cancer. All the facts show that Regorafenib is likely to be a truly effective drug, worthy of our further study.

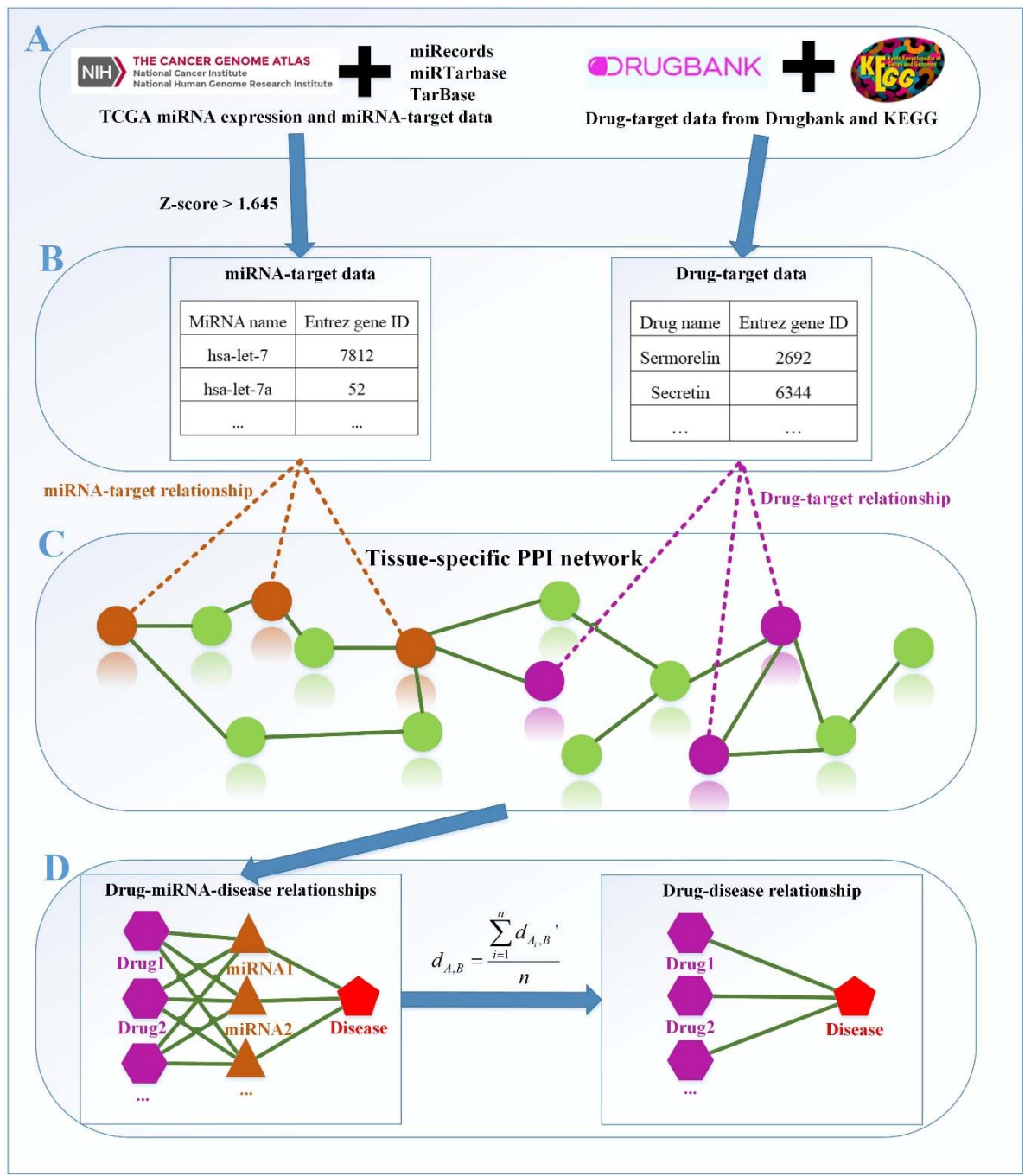

**Figure 1.** The framework of our method *miTS*. (A) Data preparation: miRNA expression data of breast cancer got from TCGA, miRNA-target gene data got from miRecords, miRTarbase and TarBase, and drug-target gene data got from Drugbank and KEGG. (B) Data preprocessing: we use Z-score to obtain the differentially expressed miRNAs for diseases and preprocess the target information of drugs. (C) In the tissue-specific PPI network, the targets of drug and miRNA are mapped to the PPI network. Orange nodes represent the target genes of miRNAs. Purple nodes represent the target genes of drugs. Green nodes represent the background genes. (D) Based on the module distance algorithm, we construct a drug-miRNA-disease tripartite network, and then based on the tripartite network, we get potential drugs for diseases. $d_{A,B}$ represents the association score between a drug and a disease.

# Data and Method

## Data

**Drug-target data:** FDA-approved drugs of human and their corresponding targets are downloaded from KEGG database and Drugbank. We merge the two datasets and get 1,732 drugs, 1,714 targets and 12,361 drug-target pairs.

**miRNA-target data:** The target genes of miRNAs are downloaded from miRecords, miRTarbase, and TarBase databases. We merge the three datasets and get 340 miRNAs, 2,028 targets and 3,652 miRNA-target pairs.

**miRNA-disease data:** The miRNA-disease curated relationships are downloaded from HMDD (the Human microRNA Disease Database)[30]. HMDD presents more detailed and comprehensive annotations to the human miRNA-disease association data, including miRNA-disease association data from the evidence of genetics, epigenetics, circulating miRNAs, and miRNA-target interactions. Finally, we get 578 miRNAs, 383 diseases and 6,448 miRNA-disease relationships.

**miRNA expression data:** Taking breast cancer as case study, we download the miRNAs expression data related with breast cancer from TCGA and get a matrix of 503 rows and 1,189 columns, row representing miRNA, column representing cancer sample, and the values in the matrix representing the RPKM (Reads Per Kilobase per Million mapped reads) for the miRNAs. We take the mean value of the RPKM values for 1,189 samples as the final value.

**Disease-gene data:** The genes related with breast cancer are downloaded from OMIM[31] database.

**Tissue-specific PPI Interaction network:** We download the mammary tissue-specific PPI network marked as "Top Edges" from GIANT (Genome-scale Integrated Analysis of gene Networks in Tissues) database[32] (http://giant.princeton.edu/) (2017 version). GIANT proposes a tissue-specific benchmark to automatically up-weight datasets relevant to a tissue from a large data of different tissues and cell-types. Finally, we get 15,269 proteins and 883,071 protein-protein interactions. The weights on the edges are proportional to the relationships between nodes. In order to apply module distance algorithm[33] to calculate the relationships between drugs and miRNAs, we use the Gaussian kernel $e^{-w^2}$ to transfer protein-protein closeness $w$ to protein-protein distance $w'$, as shown in formula (1).

$$w' = e^{-w^2} \qquad (1)$$

# Method

## Screening differentially expressed miRNAs

In order to obtain the differentially expressed miRNAs of breast cancer, we first filter the miRNAs expression data downloaded from TCGA. For a miRNA $r$, we use formula (2) to calculate its $Z\text{-}score$.

$$Z\text{-}score = \frac{R - \text{mean}(r)}{\sigma(r)} \qquad (2)$$

Where $R$ is the RPKM value of miRNA $r$; $\text{mean}(r)$ and $\sigma(r)$ represent mean value and standard deviation of $r$, respectively. Then we choose $Z\text{-}score = 1.645$ (p-value = 0.05) as threshold to screen differentially expressed miRNAs. Finally, we get a total of 40 differentially expressed miRNAs of breast cancer (see Table 1). In Table 1, the miRNAs marked by "*" represent they have connections with breast cancer in HMDD. We find 34 of 40 (85%) differentially expressed miRNAs are related with breast cancer, which indicates that miRNAs associated with breast cancer tend to be highly expressed in breast cancer patients. Then, we choose the 34 miRNAs marked by "*" in Table 1 for further study.

Table 1. Differentially expressed miRNAs of breast cancer

| miRNA name | Z-score | miRNA name | Z-score | miRNA name | Z-score |
| --- | --- | --- | --- | --- | --- |
| hsa-mir-21* | 3.32 | hsa-mir-375* | 2.32 | hsa-mir-23a* | 1.91 |
| hsa-mir-22* | 2.94 | hsa-mir-101-1* | 2.29 | hsa-mir-199a-2* | 1.90 |
| hsa-mir-10b* | 2.93 | hsa-mir-200c* | 2.28 | hsa-mir-126* | 1.90 |
| hsa-mir-30a* | 2.85 | hsa-mir-25* | 2.27 | hsa-mir-100* | 1.86 |
| hsa-mir-148a* | 2.77 | hsa-let-7a-3* | 2.21 | hsa-let-7c* | 1.79 |
| hsa-mir-99b | 2.73 | hsa-let-7a-1* | 2.21 | hsa-mir-151 | 1.78 |
| hsa-mir-143* | 2.73 | hsa-mir-30d* | 2.19 | hsa-mir-199a-1* | 1.73 |
| hsa-mir-182* | 2.72 | hsa-mir-92a-2* | 2.18 | hsa-mir-26a-2* | 1.72 |
| hsa-let-7b* | 2.61 | hsa-let-7f-2* | 2.12 | hsa-mir-142 | 1.72 |
| hsa-mir-10a* | 2.56 | hsa-mir-93* | 2.03 | hsa-mir-29c* | 1.70 |
| hsa-mir-103-1 | 2.50 | hsa-mir-29a* | 2.03 | hsa-mir-181a-1* | 1.69 |
| hsa-let-7a-2* | 2.44 | hsa-mir-28 | 2.00 | hsa-mir-141* | 1.66 |
| hsa-mir-30e | 2.38 | hsa-mir-199b* | 1.98 | | |
| hsa-mir-183* | 2.37 | hsa-mir-203* | 1.94 | | |

The miRNAs marked by "*" represent they have relationship with breast cancer in HMDD.

## Construct drug-miRNA-disease tripartite network

The relationship between a miRNA and a drug is derived by measuring the correlation between their target sets. Because miRNA target genes, drug target genes and protein-protein interaction (PPI) networks remain largely incomplete, we calculate the distance between two modules based on the shortest path in incomplete networks[33].

Figure 2 gives an example to calculate the distance between miRNA *A* and drug *B* in a weighted tissue-specific PPI network. As shown in Figure 2, miRNA *A* has three target genes, marked as *a*, *b*, *c* and drug *B* has four targets, marked as *c*, *d*, *e*, *f*. For the node *a*, its distance to targets {*c, d, e, f*} of drug *B* are 0.8, 1.0, 1.1 and 1.9 respectively, so its shortest distance to drug *B* is 0.8. In this way, we can obtain the distances between each node in gene set {*a, b, c*} and drug *B*, and the distances between each node in target set {*c, d, e, f*} and miRNA *A*, shown in Figure 2. Finally, the distance between miRNA *A* and drug *B*, $d_{A,B}'$, is equals to the sum of all the distances divided by the total number of nodes related to miRNA *A* and drug *B*. Here, the total number is 7.

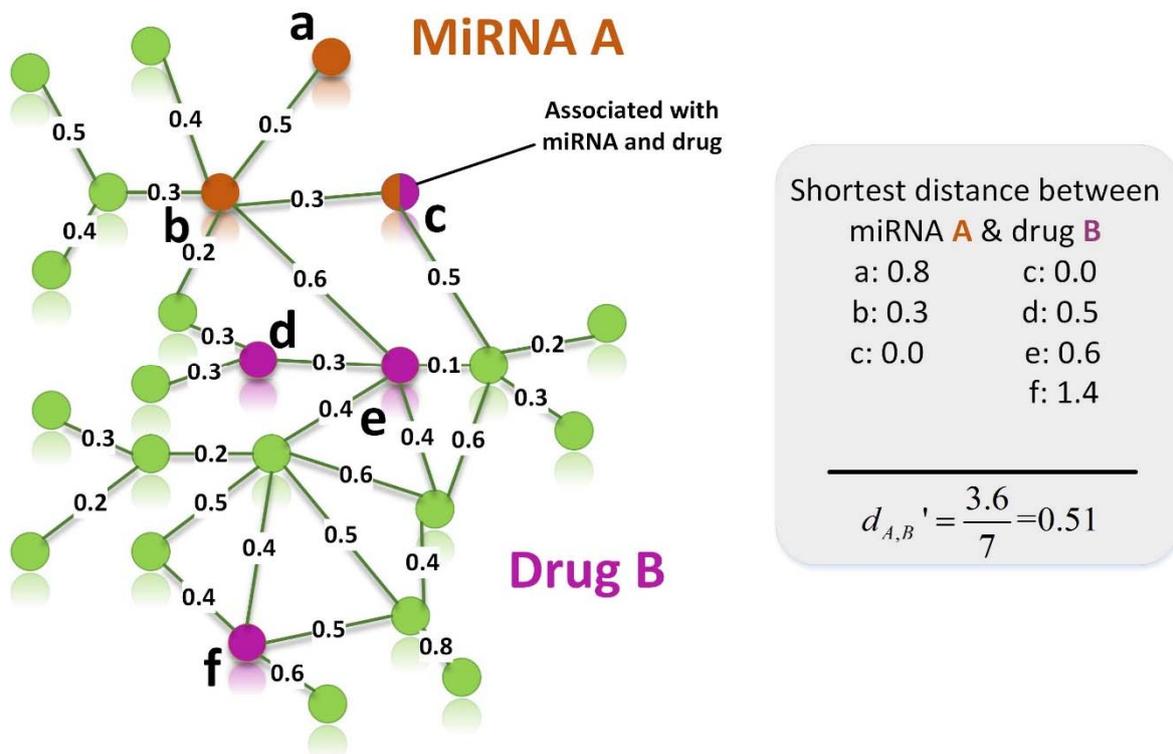

**Figure 2.** An example for calculating the distance between target set of miRNA *A* and target set of drug *B*. Orange and purple nodes represent genes related to miRNA *A* and drug *B*, respectively. Node *c* is a shared node, so it is marked by two colors.

Through the above calculation process, we get 1,017 drugs, 25 miRNAs and 25,425 drug-miRNA relationships. Combining the drug-miRNA relations with the miRNA-breast cancer information, we construct a drug-miRNA-breast cancer tripartite network.

## Predicting potential drugs for breast cancer

Based on the drug-miRNA-breast cancer tripartite network, we predicting potential drugs for breast cancer. If a drug and breast cancer have common miRNA neighbors, there will be a connection between them. Finally, there are 25 common miRNAs between drugs and breast cancer. We use formula (3) to calculate the average distance between the 25 miRNAs related to breast cancer and drugs as the drug-breast cancer relationship distance score, $d_{A,B}$.

$$d_{A,B} = \frac{\sum_{i=1}^{n} d_{A_i,B}'}{n} \quad (3)$$

Where $d_{A_i,B}'$ represents the distance between the $i$-th miRNA of disease $A$ and the drug $B$; $n$ represents the number of miRNAs corresponding to disease $A$. Here, $A =$ breast cancer and $n = 25$.

In order to make the drug-disease distances be proportional to their direct correlations, we use formula (4) to normalize $d_{A,B}$ as $S_{A,B}$:

$$S_{A,B} = \frac{Max_d - d_{A,B}}{Max_d - Min_d} \quad (4)$$

Where $Max_d$ and $Min_d$ represent the maximum and the minimum of all the drug-disease distances, respectively; $d_{A,B}$ represents the distance between disease $A$ and drug $B$; $S_{A,B}$ represents the direct association between disease $A$ and drug $B$.

# Results

## CTD benchmark verification

In our study, we choose breast cancer as case, the drug-breast cancer associations are ranked in descending order according to their scores. In order to verify the accuracy of our results, we use the drug-breast cancer relationships data in Comparative Toxicogenomics Database (CTD)[34] as benchmark. As shown in Figure 3, we give the precision curves of predicted drug-breast cancer relationship results. For each given threshold, the precision of our method is calculated by formula (5).

$$precision = \frac{P_{CTD}}{P} \qquad (5)$$

Where $P$ represents the number of predicted drug-disease pairs; $P_{CTD}$ represents the number of drug-disease pairs, which can be found in CTD database.

In Figure 3, we give the precision curves of predicted drug-breast cancer pairs at different top-x%. From the figure, we find the higher the associations ranking, the higher the accuracy. Hence, for the breast cancer, we choose top 30 drugs for further analysis. The top 30 drugs related to breast cancer are shown in Table 2. We validate the 30 drugs by CTD database and find 11 (36.7%) of them are marked as "therapeutic (T)", which means that they have a highly correlation with breast cancer. In addition, we find 14 of the rest 19 drugs also have connections with breast cancer in CTD database with inference score over 0 and they are marked as "Ref" in Table 2. That is to say, there are 83.3% (25/30) drugs can be found in the CTD database and we predict five potential drugs for breast cancer (DB08871, DB00031, DB08813, DB08896, and DB06813, marked as boldface in Table 2).

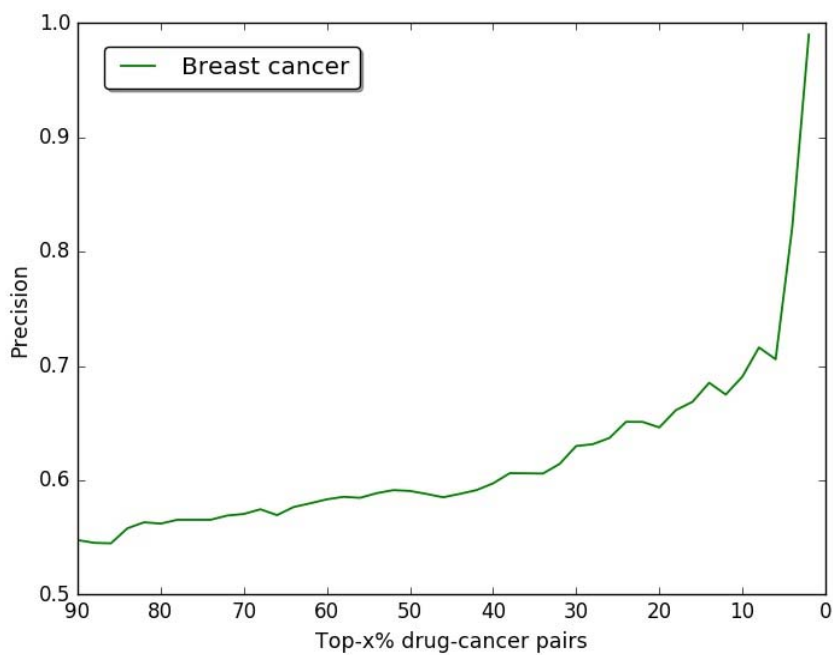

**Figure 3.** The precision of our predictions at different top-x% drug-breast cancer pairs.

## Clinical evaluation

For the five predicted drugs, we further analyze them based on the ClinicalTrials.gov (https://clinicaltrials.gov/). ClinicalTrials.gov is a registry and results database of publicly and privately supported clinical studies of human participants conducted around the world. Currently, it lists 242,537 studies with locations in all 50 states and in 198 countries (April 25, 2017). From the ClinicalTrials.gov, we can find 84 records for drug Eribulin mesylate (DB08871) treat breast cancer. For example, "Eribulin Mesylate Phase IV Clinical Trial in Korean Patients with Metastatic or Locally Advanced Breast Cancer (ESKIMO) (NCT01961544)", the purpose is assessing the safety of Eribulin which is approved for the treatment of the patients in Korea with locally advanced or metastatic breast cancer; "Eribulin with Trastuzumab as First-line Therapy for Locally Recurrent or Metastatic HER2 Positive Breast Cancer

Table 2. The top 30 drugs related to breast cancer

| Rank | Drugbank ID | Drug name | Marker | Inference Score | Similarity Score |
|------|-------------|-----------|--------|-----------------|------------------|
| 1 | DB08818 | Hyaluronic acid | Ref | 61.1 | 1.00000 |
| 2 | DB00570 | Vinblastine | T | 40.6 | 0.97977 |
| 3 | DB00642 | Pemetrexed | T | 12.38 | 0.97666 |
| 4 | DB01169 | Arsenic trioxide | T | 212.32 | 0.96998 |
| 5 | DB00242 | Cladribine | Ref | 14.83 | 0.96343 |
| 6 | DB04967 | Lucanthone | Ref | 45.07 | 0.96120 |
| 7 | DB09073 | Palbociclib | T | 53.75 | 0.96083 |
| 8 | DB02701 | Nicotinamide | Ref | 63.26 | 0.96013 |
| 9 | DB01005 | Hydroxyurea | Ref | 32.57 | 0.95909 |
| 10 | DB01204 | Mitoxantrone | T | 25.1 | 0.95847 |
| 11 | DB00309 | Vindesine sulfate | T | 2.54 | 0.95388 |
| 12 | DB00361 | Vinorelbine | T | 4.36 | 0.95388 |
| **13** | **DB08871** | **Eribulin** | **None** | **None** | **0.95388** |
| 14 | DB01394 | Colchicine | Ref | 50.29 | 0.95213 |
| 15 | DB01229 | Paclitaxel | T | 111.41 | 0.95050 |
| 16 | DB01248 | Docetaxel | T | 72.35 | 0.95050 |
| 17 | DB00440 | Trimethoprim | Ref | 6.82 | 0.94971 |
| 18 | DB01179 | Podofilox | Ref | 2.87 | 0.94918 |
| 19 | DB05260 | Gallium nitrate | Ref | 19.88 | 0.94671 |
| 20 | DB00441 | Gemcitabine | T | 112.67 | 0.94582 |
| **21** | **DB00031** | **Tenecteplase** | **None** | **None** | **0.94357** |
| **22** | **DB08813** | **Nadroparin** | **None** | **None** | **0.94325** |
| 23 | DB00432 | Trifluridine | Ref | 12.49 | 0.94205 |
| 24 | DB01073 | Fludarabine | Ref | 59.43 | 0.94184 |
| 25 | DB00694 | Daunorubicin | Ref | 85.05 | 0.94044 |
| 26 | DB00970 | Dactinomycin | Ref | 98.94 | 0.93988 |
| **27** | **DB08896** | **Regorafenib** | **None** | **None** | **0.93853** |
| **28** | **DB06813** | **Pralatrexate** | **None** | **None** | **0.93853** |
| 29 | DB00563 | Methotrexate | T | 123.36 | 0.93799 |
| 30 | DB00615 | Rifabutin | Ref | 2.89 | 0.93433 |

Ranked by drug-breast cancer similarity score. Marker has three values: T(therapeutic), Ref (inferred by genes) and None (no record in CTD database). Inference Score represents the score for the inference based on the topology of the network consisting of the chemical, disease, and one or more genes used to make the inference.

(NCT01269346)", the purpose is evaluating the safety and efficacy of Eribulin mesylate in combination with trastuzumab as first line treatment in female subjects with locally recurrent or metastatic human epidermal growth factor receptor (HER2) positive breast cancer; "Eribulin Mesylate in Treating Patients with Previously Treated Metastatic Breast Cancer (NCT01908101)", and so on. For drug Nadroparin (DB08813), we find one record: "Prevention of Venous and Arterial

Thromboembolism, in Cancer Patients Undergoing Chemotherapy, With a Low Molecular Weight Heparin (Nadroparin Calcium) (NCT00951574)", 1200 patients with lung, breast, gastrointestinal (stomach, colon-rectum, pancreas), ovarian or head and neck cancer undergoing chemotherapy will be randomly assigned in a 2:1 ratio and in double-blind conditions to a treatment with subcutaneous low molecular weight heparin (nadroparin calcium, one injection/day) or placebo for the overall duration of chemotherapy or up to a maximum of 4 months. For drug Regorafenib (DB08896), we find three records related with breast cancer, "Refametinib in Combination with Regorafenib in Patients with Advanced or Metastatic Cancer (NCT02168777)", "Effect of Regorafenib on Digoxin and Rosuvastatin in Patients with Advanced Solid Malignant Tumors (NCT02106845)", and so on. For drug Pralatrexate (DB06813), we find a clinical study of Pralatrexate in 22 female patients with previously-treated breast cancer (NCT01118624). Only one drug, Tenecteplase (DB00031) was not found in ClinicalTrials.gov.

## Literature curation

In the above section, the top 30 drugs related with breast cancer are validated by CTD database and Clinical database. After our analysis, we obtain five potential drugs (Eribulin mesylate, Tenecteplase, Nadroparin, Regorafenib, Pralatrexate) for breast cancer. In ClinicalTrials.gov database, only one drug, Tenecteplase, cannot be found its corresponding record. In this section, we will analyze the five potential drugs for breast cancer by literature mining.

Eribulin mesylate (DB08871) is an anticancer drug marketed by Eisai Co. under the trade name Halaven, which was approved by the U.S. Food and Drug Administration (FDA) on November 15, 2010, to treat patients with metastatic breast cancer[35]. In 2016, Kurebayashi J et al. investigated the combined effects of Eribulin and antiestrogens. They used a panel of eight breast cancer cell lines, including five estrogen receptors (ER)-positive and three ER-negative cell lines. The results of this study demonstrate that Eribulin had potent antitumor effects on estrogen-stimulated ER-positive breast cancer cells[36].

Nadroparin (DB08813) is an anticoagulant belonging to a class of drugs called low molecular weight heparins (LMWHs), which is used in general and orthopedic surgery to prevent thromboembolic disorders. In 2015, Sun Y et al.[37] used the MTT test to observe the effect of different concentrations of nadroparin on the growth capacity of breast cancer cells MDA-MB-231. The purpose was to study the effect of nadroparin in the migration of breast cancer cells MDA-MB-231 and its action mechanism. The results show that nadroparin can inhibit the growth capacity of breast cancer cells MDA-MB-231 and the migration and invasion of breast cancer cells MDA-MB-231. Its mechanism is to down-regulate MMP-2 and MMP-9 expressions after combining with Integrin β3.

Regorafenib (DB08896) is an oral multi-kinase inhibitor developed by Bayer which targets angiogenic, stromal and oncogenic receptor tyrosine kinase (RTK). Regorafenib has been demonstrated to increase the overall survival of patients with metastatic colorectal cancer[38]. Stalker L et al. using regorafenib in mammary tumor cell lines, the results show regorafenib may prove clinically useful in inhibiting breast cancer cell migration and metastasis[39]. Su J C et al. investigated the potential of regorafenib to suppress metastasis of triple-negative breast cancer (TNBC) cells through targeting SHP-1/p-STAT3/VEGF-A axis and found a significant correlation between cancer cell migration and SHP-1/p-STAT3/VEGF-A expression in human TNBC cells[40].

Pralatrexate (DB06813) is an anti-cancer drug. It is the first drug approved as a treatment for patients with relapsed T-cell lymphoma[41]. Pralatrexate results in increased activity of CASP3 protein, which has been found to be necessary for normal brain development as well as its typical role in apoptosis, where it is responsible for chromatin condensation and DNA fragmentation[42].

Tenecteplase (DB00031) is a tissue plasminogen activator (tPA) produced by recombinant DNA technology using an established mammalian cell line and used as a thrombolytic drug. Nielsen VG et al.[43] to study whether tissue-type plasminogen activator (tPA) in plasma obtained from patients with breast cancer, lung cancer, pancreatic cancer and colon cancer is less than that obtained from normal individuals. The results show that tissue-type plasminogen activator-induced fibrinolysis in breast cancer, lung cancer, pancreatic cancer and colon cancer patients is enhanced. Sumiyoshi K et al.[44] found that the increase in levels of plasminogen activator and type-1 plasminogen activator inhibitor in human breast cancer may play a role in tumor progression and metastasis. Although we have not found the relationship between tenecteplase (DB00031) and breast cancer through the literatures, the drug had the similar effects as nadroparin[45]. Therefore, we infer that tenecteplase is likely to have effect on breast cancer.

## KEGG pathway functional enrichment analysis

In this section, we will further make KEGG pathway enrichment analysis on five potential drugs and their associated disease. KEGG (http://www.kegg.jp/ or http://www.genome.jp/kegg/) is an encyclopedia of genes and genomes[46]. Its primary goal is to assign functional meanings to genes and genomes both at the molecular and higher levels. Thus, drugs or diseases can be associated with certain pathways through their related genes. If a drug has overlapping KEGG pathways with a disease, the drug and the disease may have great relevance. That is, the drug may treat or cause the disease by acting on the overlapping pathways.

We use DAVID[47,48] functional annotation tool for KEGG pathway enrichment analysis. DAVID provides a comprehensive set of functional annotation tools for investigators to understand biological significance of a large number of genes. For any given gene list, DAVID is able to visualize genes on BioCarta & KEGG pathway maps, identify enriched biological themes, especially GO terms, and so on. Therefore, we use DAVID to identify overlapping KEGG pathways between potential drugs and breast cancer. The *p*-value is set to be less than 0.05.

We find Nadroparin and Regorafenib have 4 and 15 overlapping KEGG pathways with breast cancer, respectively. The details are shown in Table 3. From Table 3, we can find their corresponding *p*-values are very small.

Although the drug Eribulin mesylate has not overlapping functional pathways with the breast cancer at present, it can be enriched to "hsa04540: Gap junction". In fact, protein connexin 43 (Cx43), a part of intercellular gap junctions, is frequently down-regulated in tumors[49]. Studies have demonstrated that gap junctions (GJs) composed of connexin (Cx) proteins have the potential to modulate drug chemosensitivity in multiple tumor cells[50].

Table 3. Overlapping KEGG pathways between potential drugs and breast cancer

| Drug Name | Overlapping enriched pathways | p-value |
|---|---|---|
| Nadroparin | hsa05210: Colorectal cancer | 0.00897 |
|  | hsa05161: Hepatitis B | 0.02098 |
|  | hsa05166: HTLV-I infection | 0.03704 |
|  | hsa05200: Pathways in cancer | 0.04687 |
| Regorafenib | hsa04015: Rap1 signaling pathway | 3.41E-14 |
|  | hsa04014: Ras signaling pathway | 7.18E-14 |
|  | hsa04151: PI3K-Akt signaling pathway | 3.36E-10 |
|  | hsa05230: Central carbon metabolism in cancer | 4.38E-08 |
|  | hsa05215: Prostate cancer | 2.21E-07 |
|  | hsa05200: Pathways in cancer | 1.13E-06 |
|  | hsa05218: Melanoma | 4.76E-06 |
|  | hsa05214: Glioma | 1.65E-04 |
|  | hsa05221: Acute myeloid leukemia | 0.00404 |
|  | hsa05205: Proteoglycans in cancer | 0.00433 |
|  | hsa05220: Chronic myeloid leukemia | 0.00660 |
|  | hsa04012: ErbB signaling pathway | 0.00952 |
|  | hsa05206: MicroRNAs in cancer | 0.01157 |
|  | hsa05231: Choline metabolism in cancer | 0.01269 |
|  | hsa04722: Neurotrophin signaling pathway | 0.01761 |

For drug Tenecteplase, we find one function enrichment pathway: "hsa04610: Complement and coagulation cascades". In 2016, based on the microarray data of GSE3467 from Gene Expression Omnibus(GEO) database, Yu J et al.[51] identified the differentially expressed genes (DEGs) between 9 PTC samples and 9 normal controls. The purpose was predicted key genes and pathways in papillary thyroid carcinoma.

Their results showed that the highly expressed genes in papillary thyroid carcinoma were mainly enriched on the "hsa04610: Complement and coagulation cascades" functional pathway. As for Pralatrexate, because it has only two targets: DHFR and TYMS, it has few related KEGG pathways. That is the main reason that Pralatrexate has no overlapping KEGG pathways with breast cancer at present.

## Overlapping genes between enriched KEGG pathways

To further analyze our results, for Eribulin mesylate, Tenecteplase and Pralatrexate, we calculate the common genes between enriched pathways of each drug and those of breast cancer. The more common genes, the stronger relationship between the drug and disease. The results are shown in Figure 4A-C, respectively. The purple hexagon nodes represent the enriched pathways of a drug. The light green circular nodes represent breast cancer enriched pathways. The width of edges represents the number of common genes between two pathway sets. The wider the edge, the more the number of common genes. From Figure 4, we can find the three drugs Eribulin mesylate, Tenecteplase and Pralatrexate all have strong connection with breast cancer, which further imply the three drugs are likely to be the potential treatments of breast cancer.

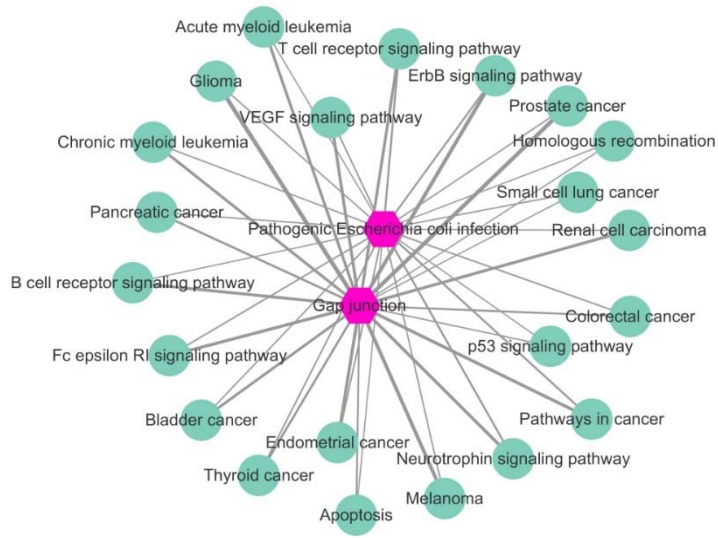

A

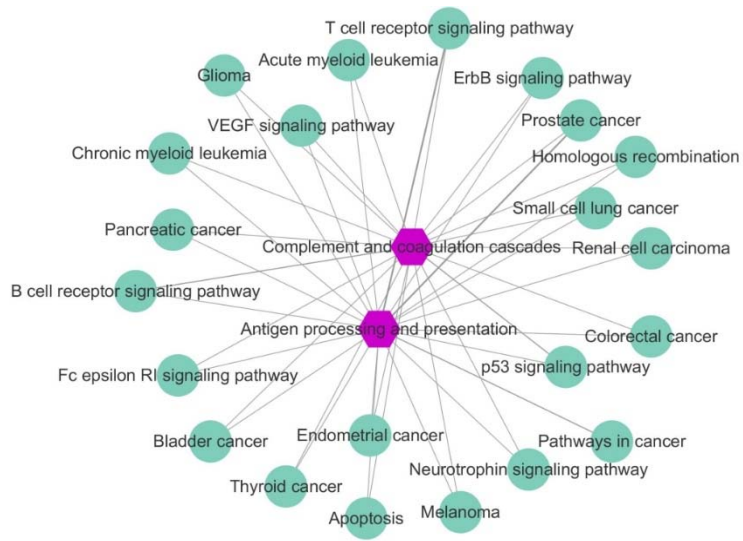

B

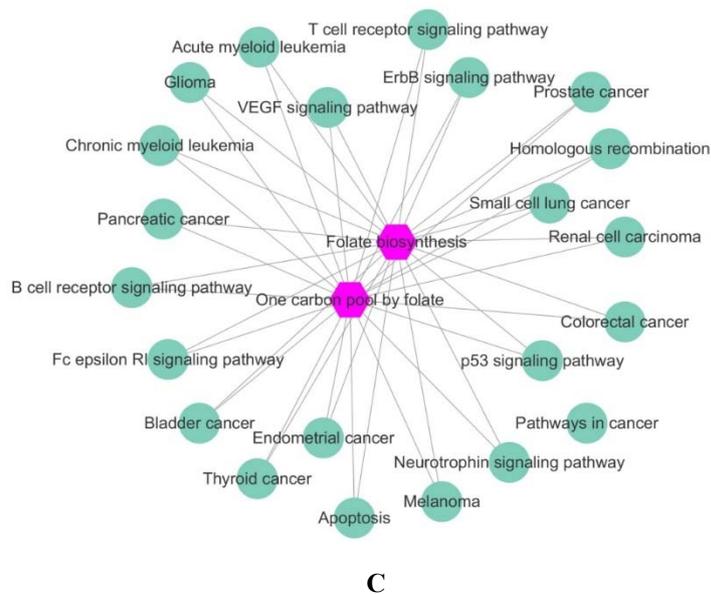

C

**Figure 4.** The common genes between enriched pathway sets of drugs and breast cancer. The purple hexagon nodes represent the enriched pathways of a drug. The light green circular nodes represent breast cancer enriched pathways. The width of edges represents the number of common genes between two pathway sets. The wider the edge, the more the number of common genes. (A) The common genes between enriched pathway sets of Eribulin mesylate and breast cancer. (B) The common genes between enriched pathway sets of Tenecteplase and breast cancer. (C) The common genes between enriched pathway sets of Pralatrexate and breast cancer.

# Discussions and conclusions

At present, "undruggable" proteins can be targeted via their miRNA gene regulators, enabling the treatment of diseases that seem impossible to cure. Human diseases result from the disordered interplay of tissue- and cell lineage–specific processes. Therefore, here we propose a new method *miTS* to predict new indications of drugs based on miRNA data and the tissue specificities of diseases. Taking breast cancer as case study, we predict five potential drugs and analyze them from five aspects: CTD benchmark, clinical records, literature curation, KEGG pathway functional enrichment analysis and overlapping genes between enriched KEGG pathways. We find for the five new drugs, they are supported at least in three ways. In particular, Regorafenib (DB08896) has 15 overlapping KEGG pathways with breast cancer and their p-values are all very small. In addition, whether in the literature curation or clinical validation, Regorafenib has a strong correlation with breast cancer. All the evidence shows Regorafenib is likely to be a truly effective drug, worthy of our further study. The results have demonstrated the performance of our model and the feasibility of drug repositioning based on miRNA data and tissue specificity.

Due to the incompleteness of data, there may be some biases in our method. With the continuous improvement of data, our method *miTS* will find more effective drugs for disease treatment. All in all, our research reveals a promising perspective to predict drug-disease relationships and seeks new opportunities for drug repositioning.

# Acknowledgments

This work was supported in part by the National Natural Science Foundation of China (Nos. 61672406, 61532014, 91530113,61502363 and 61402349), the Natural Science Basic Research Plan in Shaanxi Province of China (Nos. 2016JQ6057, 2015JM6283).